\documentclass[
    prl, twocolumn, superscriptaddress, nofootinbib, amsmath, amssymb,
    aps, floatfix
]{revtex4-2}
\usepackage{graphicx}
\usepackage{xcolor}
\usepackage{setspace}
\usepackage{hyperref}

\usepackage[utf8]{inputenc}
\usepackage[T1]{fontenc}
\usepackage{lmodern}

\usepackage{booktabs}
\usepackage{color}
\usepackage{epsfig}
\usepackage{ifpdf}
\usepackage{amsmath}
\usepackage{bm}
\usepackage[english]{babel}
\usepackage{amsfonts}
\usepackage{amssymb}

\usepackage{txfonts}

\usepackage{braket}
\usepackage{enumerate}
\usepackage{cancel}
\usepackage{multirow}
\usepackage{cleveref}
\usepackage{xspace}
\usepackage{array}
\usepackage{fancyvrb}
\usepackage{fontawesome}
\usepackage{dcolumn}
\usepackage{slashed}
\usepackage{rotating}
\usepackage[normalem]{ulem}

\newcommand{\mev}{\,\text{MeV}}
\newcommand{\gev}{\,\text{GeV}}
\newcommand{\tev}{\,\text{TeV}}
\newcommand{\invfb}{/\text{fb}}

\newcommand{\amu}{\ensuremath{a_\mu}\xspace}
\newcommand{\detaildalphahad}{\ensuremath{\Delta\alpha^{(5)}_{\textrm{had}}(M_Z^2)}\xspace}
\newcommand{\dalphahad}{\ensuremath{\Delta\alpha_{\textrm{had}}}\xspace}
\newcommand{\amuhvp}{\ensuremath{a_\mu^\text{HVP}}\xspace}
\newcommand{\sigmahad}{\ensuremath{\sigma_{\text{had}}}\xspace}

\def\Xint#1{\mathchoice
   {\XXint\displaystyle\textstyle{#1}}%
   {\XXint\textstyle\scriptstyle{#1}}%
   {\XXint\scriptstyle\scriptscriptstyle{#1}}%
   {\XXint\scriptscriptstyle\scriptscriptstyle{#1}}%
   \!\int}
\def\XXint#1#2#3{{\setbox0=\hbox{$#1{#2#3}{\int}$}
     \vcenter{\hbox{$#2#3$}}\kern-.5\wd0}}

\def\dashint{\Xint-}
\def\email#1{\hyperlink{mailto:#1}{#1}}

\definecolor{nicered}{rgb}{0.7,0.1,0.1}
\definecolor{nicegreen}{rgb}{0.1,0.5,0.1}

\begin{document}

\title{Hadronic Uncertainties versus New Physics for the $W$ boson Mass and Muon $g-2$ Anomalies}



\author{\small Peter Athron$^{1\,*}$, Andrew Fowlie$^{1\,\dagger}$, Chih-Ting Lu$^{1\,\ddagger}$, Lei Wu$^{1\,\S}$, Yongcheng Wu$^{1\,\P}$, Bin Zhu$^{2\,**}$\\
\small $^1$Department of Physics and Institute of Theoretical Physics, Nanjing Normal University,\\
\small Nanjing, 210023, China\\
\small $^2$Department of Physics, Yantai University, Yantai 264005, China\\
\small $^*$\email{peter.athron@njnu.edu.cn}\ $^\dagger$\email{andrew.j.fowlie@njnu.edu.cn}\ $^\ddagger$\email{06285@njnu.edu.cn}\\
\small $^\S$\email{leiwu@njnu.edu.cn}\ $^\P$\email{ycwu@njnu.edu.cn}\ $^{**}$\email{zhubin@mail.nankai.edu.cn}}

\begin{abstract}
There are now two single measurements of precision observables that have major anomalies in the Standard Model:
the recent CDF measurement of the $W$ mass shows a $7\sigma$ deviation and the Muon $g-2$ experiment at FNAL confirmed a long-standing anomaly, implying a $4.2 \sigma$ deviation.
Doubts regarding new physics interpretations of these anomalies could stem from uncertainties in the common hadronic contributions.
We demonstrate that these two anomalies pull the hadronic contributions in opposite directions by performing electroweak fits in which the hadronic contribution was allowed to float.
The fits show that including the $g - 2$ measurement worsens the tension with the CDF measurement and conversely that adjustments that alleviate the CDF  tension worsen the $g-2$ tension beyond $5 \sigma$.
This means that if we adopt the CDF  $W$ mass measurement, the case for new physics in either the $W$ mass or muon $g-2$ is inescapable regardless of the size of the SM hadronic contributions.
Lastly, we demonstrate that a mixed scalar leptoquark extension of the Standard Model could explain both anomalies simultaneously.
\end{abstract}

\maketitle
\newpage

\section{Introduction}\label{sec1}

The CDF collaboration at Fermilab recently reported the world's most precise direct measurement of the $W$ boson mass, $M^{\rm CDF}_W = 80.4335 \pm 0.0094\gev$~\cite{CDF:2022hxs}, based on $8.8\invfb$ of data collected between 2002-2011. This deviates from the Standard Model (SM) prediction by about $7\sigma$. The recent FNAL E989 measurement of the muon's anomalous magnetic moment furthermore implies a new world average of $\amu = 16\,592\,061(41) \times 10^{-11}$~\cite{Muong-2:2021ojo}, which is in $4.2\sigma$ tension with the SM theory prediction from the Muon $g-2$ Theory Initiative, $116\,591\,810(43)\times 10^{-11}$~\cite{Aoyama:2020ynm}.
This prediction is based on results from Refs.\ \cite{Aoyama:2012wk,Aoyama:2019ryr,Czarnecki:2002nt,Gnendiger:2013pva,Davier:2017zfy,Keshavarzi:2018mgv,Colangelo:2018mtw,Hoferichter:2019gzf,Davier:2019can,Keshavarzi:2019abf,Kurz:2014wya,Melnikov:2003xd,Masjuan:2017tvw,Colangelo:2017fiz,Hoferichter:2018kwz,Gerardin:2019vio,Bijnens:2019ghy,Colangelo:2019uex,Pauk:2014rta,Danilkin:2016hnh,Jegerlehner:2017gek,Knecht:2018sci,Eichmann:2019bqf,Roig:2019reh,Blum:2019ugy,Colangelo:2014qya}.

\begin{figure}[th]
    \centering
    \includegraphics[width=0.95\linewidth]{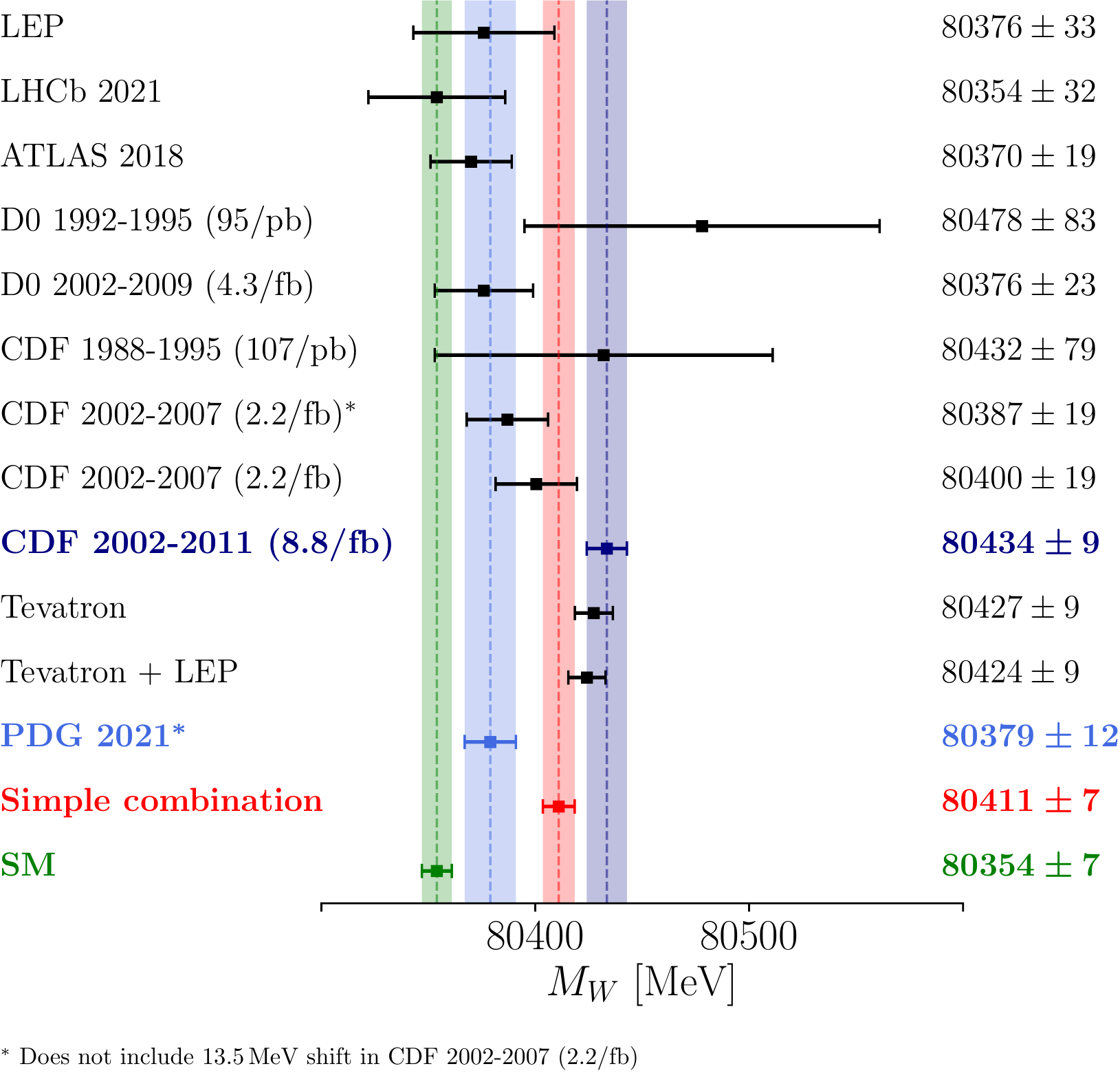}
    \caption{{\bf Simple combination of $M_W$ measured by different experiments}. Measurements of the $W$ boson mass from LEP~\cite{ALEPH:2013dgf}, LHCb~\cite{LHCb:2021bjt}, ATLAS~\cite{ATLAS:2017rzl}, D0~\cite{D0:2012kms} and CDF I and II~\cite{CDF:2012gpf,CDF:2013dpa,CDF:2022hxs}. We show an SM prediction~\cite{Haller:2018nnx}, the previous PDG combination of measurements~\cite{Zyla:2020zbs}, CDF combinations of Tevatron and LEP measurements~\cite{CDF:2022hxs}, and a simple combination that includes the new measurement, which is explained in the \protect\hyperlink{sm}{Supplementary Note 1}. The PDG combination includes uncorrected CDF measurements. The error bars show $1\sigma$ errors. The code to reproduce this figure is available at \cite{github_w_mass_combination}.}
    \label{fig:mw}
\end{figure}

Whilst the Fermilab $g-2$ measurement was in agreement with the previous BNL E821 measurement~\cite{Muong-2:2006rrc}, as shown in \cref{fig:mw} there appears to be tension between the new CDF  measurement and previous measurements, including the previous CDF  measurement with only $2.2\invfb$ of data~\cite{CDF:2012gpf}.  Updates to systematic uncertainties shift the previous measurement by $13.5\mev$, however, such that the CDF measurements are self-consistent. In the \hyperlink{sm}{Supplementary Note 1} we find a reduced chi-squared from a combination of $N = 7$ measurements of about $\chi^2 / (N-1) \simeq 3$ and a tension of about $2.5\sigma$. Nevertheless, we show that these two measurements could point towards physics beyond the SM with a common origin and, under reasonable assumptions, that the new CDF $W$ mass measurement pulls common hadronic contributions in a direction that significantly strengthens the case for new physics in muon~$g-2$.

We now turn to the SM predictions for the $W$ mass and muon $g-2$. Muon decay can be used to predict $M_W$ in the SM from the more precisely measured inputs, $G_\mu$, $M_Z$ and $\alpha$ (see e.g.~Ref.~\cite{Awramik:2003rn})
\begin{equation}
    M^2_W = M^2_Z \left\lbrace\frac{1}{2}+\sqrt{\frac{1}{4}-\frac{\pi\alpha}{\sqrt{2}G_{\mu}M^2_Z}\left(1+\Delta r\right)} \right\rbrace.
\end{equation}
The loop corrections are contained in $\Delta r$: full one-loop contributions were first calculated in Refs.~\cite{Sirlin:1980nh,Marciano:1980pb}, and the complete two-loop contributions are now available~\cite{Sirlin:1983ys,Djouadi:1987gn,Djouadi:1987di,Kniehl:1989yc,Consoli:1989fg,Halzen:1990je,Kniehl:1991gu,Barbieri:1992nz,Djouadi:1993ss,Fleischer:1993ub,Degrassi:1996mg,Degrassi:1996ps,Freitas:2000gg,Freitas:2002ja,Awramik:2002wn,Awramik:2003ee,Onishchenko:2002ve,Awramik:2002vu}. These have been augmented with leading three-loop  and leading four-loop corrections~\cite{Avdeev:1994db,Chetyrkin:1995ix,Chetyrkin:1995js,Chetyrkin:1996cf,Faisst:2003px,vanderBij:2000cg,Boughezal:2004ef,Boughezal:2006xk,Chetyrkin:2006bj,Schroder:2005db}. The state-of-the-art on-shell (OS) calculation of $M_W$ in the SM~\cite{Awramik:2003rn} updated with recent data gives
$80.356\gev$~\cite{Diessner:2019ebm},
whereas the $\overline{\text{MS}}$ scheme~\cite{Degrassi:2014sxa} result is about 6 MeV smaller when evaluated with the same input data. Direct estimates of the missing higher order corrections were a little smaller (4 MeV for OS and 3 MeV for $\overline{\text{MS}}$).

The predictions also suffer from parametric uncertainties, with the largest uncertainties coming from $m_t$ and may be around 9 MeV~\cite{Degrassi:2014sxa}, and depend on estimates of the hadronic contributions to the running of the fine structure constant, $\dalphahad \equiv \detaildalphahad$, defined at the scale $M_Z$ for five quark flavors. This is constrained by electroweak (EW) data and by measurements of the $e^+e^-\rightarrow \textit{hadrons}$ cross section (\sigmahad) through the principal value of the integral~\cite{Crivellin:2020zul}
\begin{equation}\label{eq:alphahad_sigma_had}
    \dalphahad = \frac{M_Z^2}{4 \pi^2 \alpha} \, \dashint^{\infty}_{m^2_{\pi^0}} \frac{\text{d}s}{M_Z^2 - s}\,\sigmahad(\sqrt{s}),
\end{equation}
where $m_{\pi^0}$ is the neutral pion mass. The parametric uncertainties may be estimated through global EW fits. For example, two recent global fits without any direct measurements of the $W$ boson mass predict $80.354 \pm 0.007\gev$~\cite{Haller:2018nnx} and $80.3591 \pm 0.0052\gev$~\cite{deBlas:2021wap} in the OS scheme. Lastly, the CDF collaboration quote $80.357 \pm 0.006\gev$~\cite{CDF:2022hxs}. While the precise central values and uncertainty estimates vary a little, all of these predictions differ from the new CDF measurement by about~$7 \sigma$.


Turning to muon $g-2$, the SM prediction for \amu includes hadronic vacuum polarization (HVP) and hadronic light-by-light (HLbL) contributions in addition to the QED and EW contributions that can be calculated perturbatively from first principles~\cite{Aoyama:2020ynm}. Although HVP is not the main contribution for \amu, it suffers from the largest uncertainty and it is hard to pin down its size.
The HLbL contributions in contrast have a significantly smaller uncertainty, with data-driven methods now providing the most precise estimates but with lattice QCD results that are consistent with these and which also contribute to the final result in Ref.~\cite{Aoyama:2020ynm}.
Two approaches are commonly used to extract the contributions from HVP. First, a traditional data-driven method in which the HVP contributions are determined from measurements of \sigmahad using the relationship~\cite{Lautrup:1968tdb}
\begin{equation}\label{eq:hvp_sigma_had}
\amuhvp = \frac{m^2_{\mu}}{12\pi^3}\int^{\infty}_{m^2_{\pi^0}}\frac{\text{d}s}{s} K(s) \, \sigmahad(\sqrt{s}),
\end{equation}
where $m_{\mu}$ and $m_{\pi^0}$ are the muon and neutral pion masses, respectively, and $K(s)$ is the kernel function as shown in Refs.~\cite{Lautrup:1968tdb,Achasov:2002bh}. This approach results in $\amuhvp=693.1(4.0)\times 10^{-10}$ with an uncertainty of less than $0.6\%$~\cite{Davier:2017zfy,Keshavarzi:2018mgv,Colangelo:2018mtw,Hoferichter:2019mqg,Davier:2019can,Keshavarzi:2019abf}. The second approach uses lattice QCD calculations. The recent leading-order lattice QCD calculations for HVP from the BMW collaboration significantly reduced the uncertainties and resulted in $\amuhvp =707.7(5.5) \times 10^{-10}$~\cite{Borsanyi:2020mff}. This, however, shows tension with the \sigmahad measurements method.

The $M_W$ and muon $g-2$ calculations are in fact connected by the fact that both \dalphahad and the HVP contributions can be extracted from the hadronic cross section, $\sigmahad(\sqrt{s})$, through \cref{eq:hvp_sigma_had,eq:alphahad_sigma_had}. We assume that the energy dependence of this cross-section, $g(\sqrt{s})$, is reliably known for $\sqrt{s} \ge m_{\pi^0}$~\cite{Keshavarzi:2018mgv,Davier:2019can}, but that the overall scale, \sigmahad, may be adjusted,
\begin{equation}\label{eq:sigma_s}
    \sigmahad(\sqrt{s}) = \sigmahad \, g(\sqrt{s}).
\end{equation}
This simple modification is similar to scenario~(3) in Ref.~\cite{Crivellin:2020zul}. There are of course more complicated possibilities, including increases and decreases in the hadronic cross section at different energies. Ref.~\cite{Passera:2008jk} considered these complicated possibilities to be implausible, though this is a somewhat subjective matter; see \hyperlink{sm}{Supplementary Note 2} for further discussion. Using \cref{eq:alphahad_sigma_had,eq:sigma_s} we may trade \sigmahad for \dalphahad giving $\dalphahad \propto \sigmahad$.
The HVP contributions depend on \dalphahad and conversely estimates of the HVP contributions from either hadronic cross-sections or lattice QCD constrain \dalphahad.
Further details of the transformation between \dalphahad and \amuhvp are provided in the \hyperlink{sm}{Supplementary Note 2}.
Thus we can transfer constraints on \dalphahad from measurements of $M_W$ to constraints on the HVP contributions to muon $g-2$ and vice-versa~\cite{Passera:2008jk,Keshavarzi:2020bfy,deRafael:2020uif} through global EW fits.

In this work, we study how the new $M_W$ measurement from CDF impacts estimates of muon $g-2$ in global EW fits and show that a common explanation of muon $g-2$ and the CDF $M_W$ from hadronic uncertainties are not possible. Then we demonstrate that in contrast a scalar leptoquark model could provide a simultaneous explanation of both muon $g-2$ and the $W$ mass anomalies.



\section{Results and discussion}

\subsection{Electroweak Fits of the $W$ mass and Muon $g-2$}\label{sec2}

\begin{figure}[t]
    \includegraphics[width=8cm]{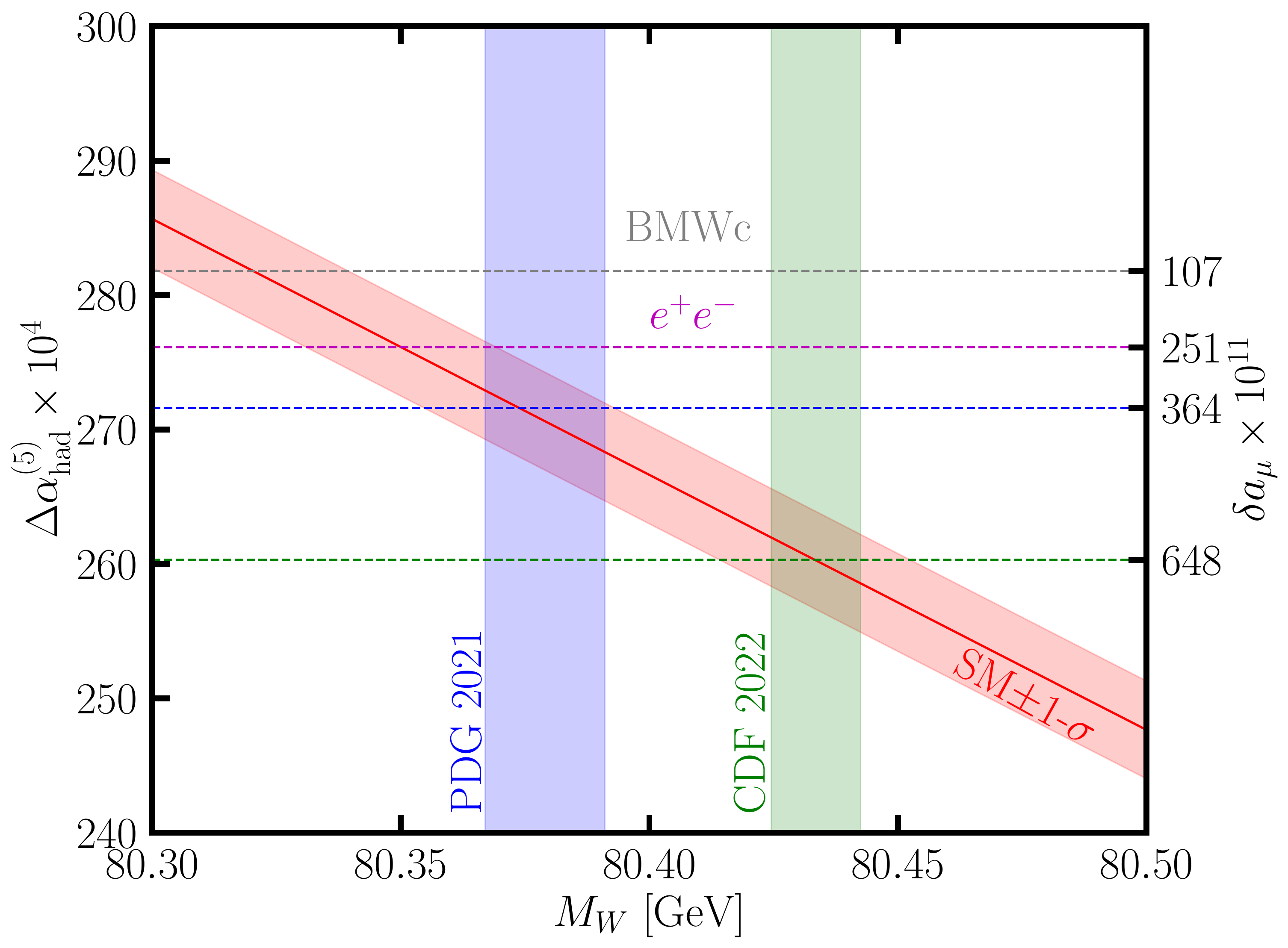}
    \caption{{\bf Correlation of \dalphahad with $M_W$ in the electroweak fits.} The allowed values of \dalphahad assuming specific values of the $W$ boson mass in the SM found from EW fits. The solid line indicates the central value from the fit without any input for \dalphahad, while the red band shows the $1\sigma$ region. The current world average (PDG 2021) and new measurement (CDF 2022) for $M_W$ are indicated by vertical bands. We indicate the favored \dalphahad from BMW lattice calculations (gray), $e^+e^-$ cross section measurements (magenta), our fit using $M_W$ from PDG 2021 (blue) and our fit using $M_W$ from CDF 2022 (green).}
    \label{fig:mw_mt_dalphahad}
\end{figure}

We first investigated the impact of the $W$ mass on the allowed values of \dalphahad by performing EW fits using \texttt{Gfitter}~\cite{Flacher:2008zq,Baak:2011ze,Baak:2012kk,Baak:2014ora,Haller:2018nnx} with data shown in~\hyperlink{sm}{Supplimentary Table 1} where $m_h$, $m_t$, $M_Z$, $\alpha_s$ and \dalphahad were allowed to float. The Fermi constant $G_F = 1.1663787\times 10^{-5}\,\text{GeV}^{-2}$ and the fine-structure constant $\alpha=1/137.035999074$~\cite{ParticleDataGroup:2020ssz} in the Thompson limit were fixed in our calculation. Although \dalphahad is not a free parameter of the SM as it is in principle calculable, it isn't precisely known and so we allowed it to float, following the approach used in Ref.\ \cite{Crivellin:2020zul}.
We found the allowed $\dalphahad$ when assuming specific $W$ masses between $80.3\gev$ and $80.5\gev$; the results form the diagonal red band in~\cref{fig:mw_mt_dalphahad}.
We fixed $\Delta M_W = 9.4$ MeV when obtaining the $\pm1\sigma$ region.
The previous world average (PDG 2021) and current CDF measurement (CDF 2022) of the $W$ mass are shown by blue and green vertical bands, respectively, and the corresponding best-fit \dalphahad are indicated by blue and green dashed horizontal lines, respectively.
From the intersection of regions allowed by CDF 2022 (green) and the EW fit (red), we see that the CDF measurement pulls \dalphahad down to about $260 \times 10^{-4}$, making the muon $g-2$ discrepancy even worse. Indeed, unless the CDF measurement is entirely disregarded it must increase the tension between the muon $g-2$ measurements and the SM prediction. The overall best-fits were found at around $M_W \simeq 80.35\gev$, in agreement with previously published fits.

\begin{table*}[th]
    \centering
    \resizebox*{\textwidth}{!}{
    \renewcommand{\arraystretch}{1.25}
    \begin{tabular}{c|c|ccc|ccc|ccc|ccc}
    \hline\hline
    \multicolumn{2}{c|}{$M_W$}& \multicolumn{3}{c|}{Indirect} & \multicolumn{3}{c|}{PDG 2021} & \multicolumn{3}{c|}{CDF 2022}  & \multicolumn{3}{c}{Simple Combination} \\
    \multicolumn{2}{c|}{\dalphahad}& BMWc & $e^+e^-$ & Indirect & BMWc & $e^+e^-$ & Indirect & BMWc & $e^+e^-$ & Indirect & BMWc & $e^+e^-$ & Indirect \\
    \hline
    \multirow{2}{*}{Input} & $M_W$ [GeV] & - & - & - & 80.379(12) & 80.379(12) & 80.379(12) & 80.4335(94) & 80.4335(94) & 80.4335(94) & 80.411(7) & 80.411(7) & 80.411(7) \\
    & $\dalphahad\times10^4$ & 281.8(1.5) & 276.1(1.1) & - & 281.8(1.5) & 276.1(1.1) & - & 281.8(1.5) & 276.1(1.1) & - & 281.8(1.5) & 276.1(1.1) & - \\
    \hline
    \multirow{7}{*}{Fitted}
    & $\chi^2/\rm dof$ & 18.32/15 & 16.01/15 & 15.89/14 & 23.41/16 & 18.74/16 & 17.59/15 & 74.51/16 & 62.58/16 & 47.19/15 & 62.08/16 & 49.79/16 & 35.48/15 \\
    \cline{2-14}
    &$M_W$ [GeV] & 80.348(6)  &  80.357(6) &  80.359(9) & 80.355(6) & 80.361(6) & 80.367(7) & 80.375(5) & 80.380(5) & 80.396(7) & 80.377(5) & 80.381(5) & 80.393(6) \\
    & $\dalphahad \times10^4$ &  280.9(1.4) & 275.9(1.1) & 274.4(4.4) & 280.3(1.4) & 275.6(1.1) & 271.7(3.8) & 278.6(1.4)  & 274.7(1.0) & 260.9(3.6) & 278.5(1.4) & 274.6(1.0) & 262.3(3.4) \\
    \cline{2-14}
    & $\delta a_\mu\times10^{11}$ & - & - & 294(166) & 146(68) & 264(59) & 364(145) & 188(68) & 289(57) & 648(137) & 191(68) & 289(57) & 597(132) \\
    & Tension & - & - & $1.8\sigma$ & $2.1\sigma$ & $4.5\sigma$ & $2.5\sigma$ & $2.8\sigma$ & $5.1\sigma$ & $4.7\sigma$ & $2.8\sigma$ & $5.1\sigma$ & $4.5\sigma$ \\
    \cline{2-14}
    & $\delta M_W$ [MeV] & 86(11) & 77(11) & 75(13) & 79(11) & 73(11) & 67(12) & 59(11) & 54(11) & 38(12) & 57(11) & 53(11) & 41(11) \\
    & Tension & $7.8\sigma$ & $7.0\sigma$ & $5.8\sigma$ & $7.2\sigma$ & $6.6\sigma$ & $5.6\sigma$ & $5.4\sigma$ & $4.9\sigma$ & $3.2\sigma$ & $5.2\sigma$ & $4.8\sigma$ & $3.7\sigma$ \\
    \hline\hline
    \end{tabular}
    }
    \caption{SM predictions from EW fits for \dalphahad and $M_W$, and the differences with respect to measurements of muon $g-2$ and the $W$ mass, $\delta \amu$ and $\delta M_W \equiv M_W^{\text{CDF}} - M_W$. The input data for \dalphahad and $M_W$ are listed in first two rows for each case.
    \label{tab:fits}
    }
\end{table*}

We further scrutinize the impact of assumptions about the HVP contributions and the $W$ mass through several fits shown in~\cref{tab:fits}.
In the first three fits, the $W$ mass is only indirectly constrained by EW data, and \dalphahad is constrained by the BMWc determination of the HVP contributions, by the $e^+e^-$ data, and indirectly by EW data. The second and final three fits are similar, though the $W$ mass is constrained by the PDG 2021 world average and by the CDF 2022 measurement, respectively. In each case we show the overall goodness of fit, and how much the best-fit muon $g-2$ and $W$ mass predictions deviate from the world average and the recent CDF measurement, respectively. Regardless of the constraints imposed on \dalphahad, including the CDF measurement results in poor overall goodness of fit and increases the tension between the SM prediction for $g-2$ and the world average. The tension between the SM prediction for the $W$ mass and the CDF measurement range from $3.2\sigma$ to $7.8\sigma$. However, the former occurs only when estimates of HVP are completely ignored (final column) and at the expense of increased tension in the SM $g-2$ prediction and a poor overall goodness of fit. Note that this includes the scenario where we do not include any input values for $M_W$ or \dalphahad in the global EW fit, as shown in the third data column (of twelve). Even in this case there is still a large tension with the CDF measurement ($5.8\sigma$), indicating that other EW observables also constrain \dalphahad. Using the $e^+e^-$ estimates of HVP, which is a standard choice, we see about $5\sigma$ tension in both $g-2$ and the $W$ mass. In fact, the CDF measurement takes the tension between the SM prediction for muon $g-2$ and the measurements slightly beyond $5\sigma$. Switching to BMWc estimates of HVP  partially alleviates the tension in $g-2$  but results in increased tension with the CDF $W$ mass measurement.

In summary, our fits showed the extent to which the new $W$ mass measurement worsens tension with muon $g-2$, using the reasonable assumption that the energy-dependence of the hadronic cross section that connects these is well-known and not modified by for example very light new physics. The anomalies pull \dalphahad in opposite directions in EW fits, making it even harder to explain both within the SM. We thus now turn to a new physics explanation.

\subsection{Interpretation in Scalar Leptoquark Model}\label{sec3}

Even without light new physics, sizable BSM contributions to muon $g-2$ can be obtained by an operator that gives an internal chirality flip in the one-loop muon $g-2$ corrections (see e.g. Refs.~\cite{Stockinger:2006zn,Athron:2021iuf} for a review). On the other hand, BSM contributions to the $W$ mass can be obtained when there are large corrections to the oblique parameter $T$~\cite{Peskin:1991sw}.  We show that a scalar leptoquark model can satisfy both of these criteria and provide a simultaneous explanation of both muon $g-2$ and the $W$ mass anomalies. We anticipate other possibilities, including composite models with non-standard Higgs bosons~\cite{Cacciapaglia:2022xih}.

Scalar leptoquarks (LQs) (see Ref.\ \cite{Dorsner:2016wpm} for a review), or more specifically the scalar leptoquarks referred to as $S_1$ $({\boldmath \overline{3}}, {\boldmath 1}, 1/3)$  and $R_2$ $({\boldmath 3}, {\boldmath 2}, 7/6)$ in Ref.\ \cite{Buchmuller:1986zs,Choi:2018stw,Lee:2021jdr}, are well known to provide the chirality flip needed to give a large contribution to \amu \cite{Chakraverty:2001yg}, and have also been proposed for a simultaneous explanation of the flavour anomalies \cite{Bauer2016}.
Furthermore due to the mass splitting between its physical states the $\text{SU}(2)$ doublet $R_2$ is capable of making a considerable contribution to the $W$ mass. However we find that the mass splitting from a conventional Higgs portal interaction cannot generate corrections big enough to reach the CDF measurement, unless the interaction $\lambda_{R_2 H} R^{\dagger}_2 H H^{\dagger} R_2$ is non-perturbative.
We thus analyze the plausibility of situations in which one-loop contributions to the anomalous muon magnetic moment and $W$ mass corrections are created via the mixing of two scalar LQs through the Higgs portal. For simplicity, we consider the $S_{1} \& S_{3}$ $({\boldmath \overline{3}}, {\boldmath 3}, 1/3)$ scenario,
\begin{equation}
\begin{aligned}
\mathcal{L}_{S_1\&S_3}=\mathcal{L}_{\mathrm{mix}}+\mathcal{L}_{\mathrm{LQ}},
\end{aligned}
\end{equation}
where the first term is responsible for the mixing of the two LQs, and the second specifies the interaction between quarks and leptons
\begin{align}
    \mathcal{L}_{\mathrm{mix }}&=\lambda H^{\dagger}\left(\vec{\tau} \cdot \overrightarrow{S_{3}}\right) H S_{1}^{*}+\text { h.c. }\\
    \mathcal{L}_{\mathrm{LQ}}&=y_{R}^{i j} \bar{u}_{R i}^{C} e_{R j} S_{1}+y_{L}^{i j} \bar{Q}_{i}^{C} i \tau_{2}\left(\vec{\tau} \cdot \vec{S}_{3}\right) L_{j}+\text{h.c.}
\end{align}
Although a coupling between $S_1$ and the left-handed lepton and quark fields is also allowed, we do not initially consider it here. Instead we show that it is possible to have new physics explanations of the CDF 2022 measurement and the 2021 combined \amu world average that originate from the same feature of the model, namely the combination of the $S_1$ and $S_3$ states through a non-vanishing mixing parameter, $\lambda$. For simplicity, we assume that only the couplings to muons that give the large chirality flipping enhancement from muon $g-2$ i.e., $y_R^{t\mu}$ and $y_L^{b\mu}$ are non-vanishing in the new Yukawa coupling.

After EW symmetry breaking, we have four scalar LQs, one with an electromagnetic charge $Q=4/3$, one with $Q = -2/3$ and two with $Q=1/3$. The $Q = 1/3$ states mix through the $\lambda$ interaction resulting in mass eigenstates $S_{+}^{\pm1/3}$ and $S_{-}^{\pm1/3}$ with masses $m_{S_+}$ and $m_{S_-}$:
\begin{align}
    \begin{pmatrix}
    S_1^{\pm1/3}\\
    S_3^{\pm1/3}
    \end{pmatrix} = \begin{pmatrix}
    c_\phi & s_\phi \\
    -s_\phi & c_\phi
    \end{pmatrix}\begin{pmatrix}
    S_+^{\pm 1/3}\\
    S_-^{\pm 1/3}
    \end{pmatrix}.
\end{align}
where $\phi$ is the mixing angle. The masses $m_{S_3}$, $m_{S_1}$ and the mixing parameter $\lambda$ can be obtained from $m_{S_+}$, $m_{S_-}$ and $\phi$ from
\begin{align}
    \delta = \frac{\lambda v^2}{2} &= s_\phi c_\phi (m_{S_-}^2-m_{S_+}^2)\\
    m_{S_1}^2 &= m_{S_+}^2c_\phi^2 + m_{S_-}^2s_\phi^2 \\
    m_{S_3}^2 &= m_{S_+}^2s_\phi^2 + m_{S_-}^2c_\phi^2
\end{align}
where $v=246$ GeV is the vacuum expectation value. We also define $\Delta m \equiv m_{S_+}-m_{S_-}$ as the mass splitting between the two mass eigenstates $S_+$ and $S_-$. This mass splitting generates a non-vanishing oblique correction to the $T$ parameter at one-loop~\cite{Dorsner:2019itg},
\begin{equation}
\begin{split}
T =\frac{3}{4 \pi s_{W}^{2}} \frac{1}{M_{W}^{2}}\Big[ & F\left(m_{S_{3}}, m_{S_{-}}\right) \cos ^{2} \phi  \\
& + F\left(m_{S_{3}}, m_{S_{+}}\right) \sin ^{2} \phi \Big]
\end{split}
\label{Eq:T_SLQs}
\end{equation}
with
\begin{equation}
F\left(m_{1}, m_{2}\right)=m_{1}^{2}+m_{2}^{2}-\frac{2 m_{1}^{2} m_{2}^{2}}{m_{1}^{2}-m_{2}^{2}} \log \left(\frac{m_{1}^{2}}{m_{2}^{2}}\right).
\end{equation}
This function vanishes when the masses are degenerate, that is, $\lim_{m_1 \to m_2} F(m_1, m_2) = 0$. When $\Delta m = 0$, the custodial symmetry is restored, and the corrections to the $T$ parameter vanish as $m_{S_3} = m_+ = m_-$.
The shift in $M_W$ from the SM prediction can be related to the oblique $T$ parameter via,
\begin{equation}
 \Delta M_{W}^{2} \equiv \left.M_W^{2}\right|_\text{BSM} - \left.M_W^{2}\right|_\text{SM} = \frac{\alpha c_W^{4} M_{Z}^{2}}{c_W^{2}-s_W^{2}} T
\label{eqn:Wmass}
\end{equation}
where $c_W$ and $s_W$ are the cosine and sine of the Weinberg angle. There are, furthermore, contributions from $S$ and $U$ that are subdominant in our LQ model.    We determine the $T$ that is required to explain the CDF 2022 measurement from our EW global fits
and use that in combination with \cref{Eq:T_SLQs} to test if LQ scenarios can explain this data. We checked analytically and numerically that our calculation obeys decoupling, with the additional BSM contributions approaching zero in the limit of large LQ masses.  We cross-checked \cref{Eq:T_SLQs} with a full one-loop calculation of the $T$ parameter using \texttt{SARAH 4.14.3} \cite{Staub:2013tta}, \texttt{FeynArts 3.11} \cite{Hahn:2000kx}, \texttt{FormCalc 9.9} \cite{Hahn:2016ebn} and \texttt{LoopTools 2.16} \cite{Hahn:1998yk}, finding good agreement with the results using just \cref{Eq:T_SLQs}. With the same setup we also verified that the combined contributions from $S$ and $U$ to $M_W$ are small and do not impact significantly on our results. Finally we also implemented this model in \texttt{FlexibleSUSY}~\cite{Athron:2014yba,Athron:2017fvs,Athron:2021kve,Athron:2022isz} using the same \texttt{SARAH} model file and the recently updated $M_W$ calculation \cite{Athron:2022isz} and again found reasonable agreement with the results of our analysis described above.

Whilst the mass splitting impacts the $W$ mass, the mixing impacts muon $g-2$. Indeed, the mixing between interaction eigenstates allows the physical mass eigenstates to have both left- and right-handed couplings to muons and induces chirality flipping enhancements to muon $g-2$ \cite{Dorsner:2019itg}
\begin{equation}
\delta \amu=-\frac{3 m_{\mu}^{2}}{16 \pi^{2}} \frac{m_{t}}{m_{\mu}} \sin2\phi \, y_L y_R \left(\frac{G(x_t^+)}{m_{S^+}^2}-\frac{G(x_t^-)}{m_{S^-}^2}\right)
\label{Eq:amu_LQ}
\end{equation}
where $x_{t}^{\pm}=m_{t}^{2} / m_{S_{\pm}}^{2}$, the loop function is ${G}(x)=1/3 g_{S}(x)-g_{F}(x)$ with
\begin{equation}
\begin{aligned}
g_{S}(x)&=\frac{1}{x-1}-\frac{\log x}{(x-1)^{2}} \\
g_{F}(x)&=\frac{x-3}{2(x-1)^{2}}+\frac{\log x}{(x-1)^{3}},
\end{aligned}
\end{equation}
and we simplify our notation by letting $y_L \equiv y_L^{b\mu}$ and $y_R \equiv y_R^{t\mu}$.
Note that in this case there is a cancellation between the contribution of the lighter and heavier mass eigenstates, which reduces the effect of the very large chirality flipping enhancement $m_t/m_\mu$ somewhat.  If we consider couplings between $S_1$ state and left-handed muons as well, the contributions would be considerably enhanced, so this would simply make it easier to explain \amu while having little or no impact on the $W$ mass prediction.

\begin{figure}[t]
    \centering
    \includegraphics[width=0.95\linewidth]{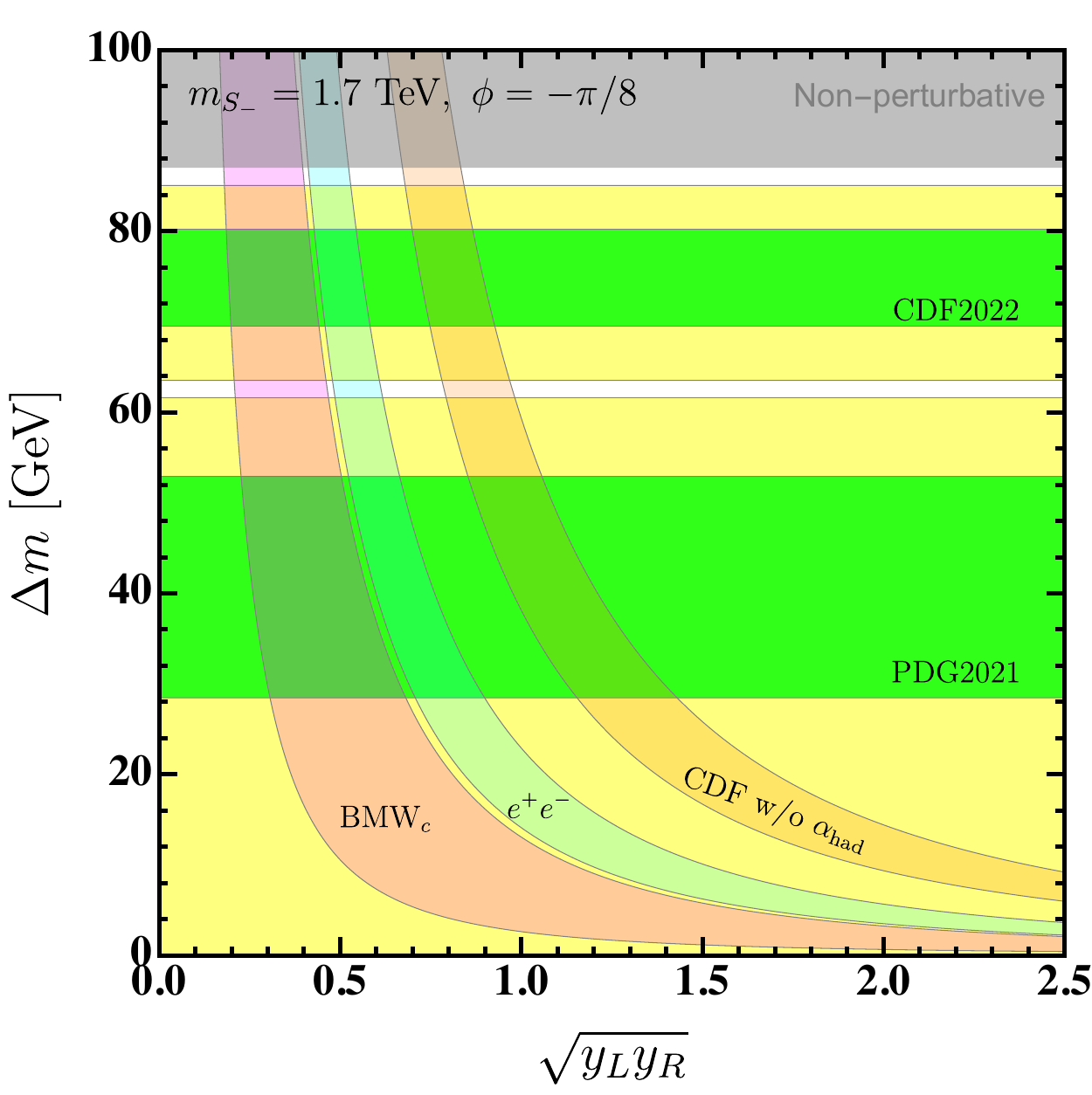}
    \caption{ {\bf Scalar leptoquark explanation.} Regions in the $\Delta m$--$\sqrt{y_L y_R}$ plane of our LQ model that predict the $W$-boson mass and muon $g-2$ in agreement with measurements. The mixing angle $\phi$ is set to be $-\pi/8$ and $m_{S_-}=1.7\,\mathrm{TeV}$. The gray region is excluded by the perturbativity requirement $|\lambda| < \sqrt{4\pi}$.}
    \label{fig:leptoquark}
\end{figure}

BSM contributions to \amu and the $W$ mass both require non-vanishing $\Delta m$. For \amu, it further requires non-vanishing mixing $\phi$ and relies on $y_Ly_R$. Thus it is possible to find explanations of both \amu and the 2022 CDF measurement of $M_W$ by varying $y_Ly_R$ and $\Delta m$ with non-zero mixing angle $\phi$.
In \cref{fig:leptoquark} we show regions in the $\Delta m$--$\sqrt{y_L y_R}$ plane that explain both measurements,
where we have fixed the LQ mass to $1.7\tev$, a little above the LHC limit, and we also fixed the mixing angle $\phi=-\pi/8$.

LQ couplings of greater than about a half can explain the \amu measurement within $1\sigma$ when we use the SM prediction from the theory white paper, where $e^+e^-$ data is used for \amuhvp.  Explaining the SM prediction from the BMW collaboration requires even smaller couplings, though in this case the tension with the SM is anyway less than $2\sigma$.  Using $e^+e^-$ data to also fix  $\dalphahad$ means there then remains an additional deviation between the SM $M_W$ prediction and the measured values. To explain the new 2022 CDF result with BSM contributions as well, $\Delta m \approx 75\gev$ is then required, and the dual explanation of the $M_W$ and \amu anomalies can be achieved in the region where the green CDF 2022 band overlaps with the light blue $e^+e^-$ band in \cref{fig:leptoquark}.  The deviation between the SM prediction for the $W$ boson mass and the 2021 PDG value is not so large and within $2\sigma$ it does not need new physics contributions, so the yellow $2\sigma$ band for this in \cref{fig:leptoquark} can extend to $\Delta m \approx 0$, but to within $1\sigma$ a small non-zero $\Delta m$ is required. Further, the interaction coupling $\lambda$ is proportional to the mass splitting $\Delta m$ with fixed mixing angle $\phi$. In order to keep the coupling perturbative ($|\lambda| < \sqrt{4\pi}$), there is an upper limit on the mass splitting as shown by the grey band in \cref{fig:leptoquark}. Note that the region that can accommodate the CDF measurement is close to the non-perturbative region, as the CDF measurement requires large mass splitting. However, it is still possible to explain the new $M_W$ measurement within $1\sigma$ in the perturbative region.

\hypertarget{sec:further_constraints}{\subsection{Further constraints}}

This model establishes a proof of principle of a simple, dual explanation of both anomalies. There remains, however, the question of whether this model or extensions of it can simultaneously explain recent flavour physics measurements and anomalies and satisfy additional phenomenological constraints. The latter may be particularly severe as the required Yukawa couplings are $\mathcal{O}(1)$.

For example, the recently measured branching ratio $\text{BR}(h\rightarrow \mu\mu)$ \cite{ATLAS:2020fzp,CMS:2020xwi} can be an important probe of leptoquark explanations of muon $g-2$~\cite{Crivellin:2020tsz,Dermisek:2022aec}. Ref.~\cite{Crivellin:2020tsz} showed that when you have $S_1$ and $S_3$ with only right-handed couplings for $S_1$ there is already a significant tension with the current measurements. There are several ways to avoid this tension. If the leptoquarks
are embedded in a more fundamental theory there could be additional light states that result in cancellations with the leptoquark contribution to $h\rightarrow\mu\mu$, for example through destructive interference between tree- and loop-level diagrams. 
This can be achieved by extending the LQ model in the framework of the two-Higgs-doublet model in the wrong-sign Yukawa coupling region~\cite{Su:2019ibd}.
Alternatively we can reintroduce the left-handed coupling of the $S_1$ state, which brings two benefits.

First, allowing significant left-handed couplings from $S_1$ substantially reduces the size of the Yukawa couplings needed to explain muon $g-2$ (as stated earlier). We show in the \hyperlink{sm}{Supplementary Note 3} that, as can be anticipated from Ref.\ \cite{Crivellin:2020tsz}, this makes it possible to satisfy the $\text{BR}(h\rightarrow \mu\mu)$ data while simultaneously explaining muon $g-2$, while keeping the mass splitting fixed to a value required to explain the CDF measurement of the $W$ mass. Second introducing this coupling gives additional freedom that can help explain the well-known anomalies of Lepton Flavour Universality Violation, while avoiding other limits from flavour physics.

Indeed, the severe constraints from $\mu\rightarrow e\gamma$, $a_e$, etc.~can all be evaded by allowing the proper flavour ansatz for the Yukawa couplings~\cite{Bhaskar:2022vgk}. At the same time the $R_{K^{\ast}}$ and $R_{D^{\ast}}$ anomalies can be explained via extra tree-level processes from the scalar LQ with $Q=4/3$ to $b\rightarrow s\mu^{+}\mu^{-}$ and two scalar LQs with $Q=1/3$ to $b\rightarrow c\tau\nu$. Although the latter two scalar LQs also contribute to $R^{\nu\nu}_{K^{\ast}}$ through tree-level process $b\rightarrow s\nu\nu$, these enhancements are under control and their effects are not in conflict with the current measurement~\cite{Bhaskar:2022vgk}.

Finally, we show that an explanation of the $W$ mass in our model must be accompanied by new physics below about $2\tev$. In order to explain the CDF measurement at $2\sigma$ with $S=0$ (which is a good approximation in our model), we need $0.14 \lesssim T \lesssim 0.17$. Expanding \cref{Eq:T_SLQs},
\begin{equation}
    T \simeq \frac{\lambda^2 v^4}{16\pi m_-^2 M_W^2 s_W^2}
\end{equation}
such that combining the lower limit on the $T$ parameter and the perturbativity limit $\lambda \le \sqrt{4\pi}$ we obtain
\begin{equation}
    m_- \lesssim 2\tev
\end{equation}
As the mass splitting, $\Delta m = m_+ - m_- \lesssim 100 \,\text{GeV}$, our model thus predicts two $Q=1/3$ LQ states below about~$2\tev$.

\section{Data availability}
The datasets generated during and/or analysed during the current study are available from the corresponding author on request.

\section{Code availability}

The custom computer codes used to generate results are available from the corresponding author on request.

\section{References}
\bibliography{refs}




\section{Acknowledgments}
\label{sec5}
We thank Martin Hoferichter for his useful comments. PA thanks Dominik St\"ockinger for helpful early discussions regarding this project. LW and BZ are supported by the National Natural Science Foundation of China (NNSFC) under grants No. 12275134 and 12275232, respectively. PA and AF are supported by the National Natural Science Foundation of China (NNSFC) Research Fund for International Excellent Young Scientists grants 12150610460 and 1950410509, respectively. Y.W. would also like to thank U.S.~Department of Energy for the financial support, under grant number DE-SC 0016013.

\section{Author Contributions Statement}

PA contributed to understanding the connections between muon $g-2$ and $M_W$, the scalar leptoquark calculations and constraints, interpreting results and to the writing of all section of the paper.
AF contributed to the introduction, statistical analysis, and interpretation and writing throughout.
CL contributed to the original idea, introduction muon $g-2$ calculation and interpretation as well writing throughout.
LW contributed to the introduction, calculations, interpretation, and writing throughout.
YW contributed to the electroweak global fits, the calculations in the leptoquark model and writing the corresponding sections.
BZ has offered a leptoquark explanation for the newly measured $W$-mass in CDF and muon $g-2$ anomaly in Fermi-Lab. He additionally examined correlation between the decay of Higgs bosons into muons and the muon $g-2$ deviation, where the dangerous constraint is removed by introducing a left-handed contribution.

\section{Competing Interests Statement}
The authors declare no competing interests.

\onecolumngrid
\clearpage


\newcommand{\ssection}[1]{
    \addtocounter{section}{1}
    \section{\thesection.~~~#1}
    \addtocounter{section}{-1}
    \refstepcounter{section}
}
\newcommand{\ssubsection}[1]{
    \addtocounter{subsection}{1}
    \subsection{\thesubsection.~~~#1}
    \addtocounter{subsection}{-1}
    \refstepcounter{subsection}
}
\newcommand{\fakeaffil}[2]{$^{#1}$\textit{#2}\\}

\thispagestyle{empty}
\begin{center}
    \begin{spacing}{1.2}
        \textbf{\large
            \hypertarget{sm}{Supplemental Material:} \boldmath Hadronic Uncertainties versus New Physics for the $W$ boson Mass and Muon $g-2$ Anomalies\\
        }
    \end{spacing}
    \par\smallskip
    Peter Athron,$^{1}$
    Andrew Fowlie,$^{1}$
    Chih-Ting Lu,$^{1}$
    Lei Wu,$^{1}$
    Yongcheng Wu,$^{1}$
    and Bin Zhu,$^{2}$
    \par
    {\small
        \fakeaffil{1}{Department of Physics and Institute of Theoretical Physics, Nanjing Normal University, Nanjing, 210023, China}
        \fakeaffil{2}{Department of Physics, Yantai University, Yantai 264005, China}
    }

\end{center}
\par\smallskip

In this supplemental material, we present the data in our global fits (\autoref{tab:fit_input}), our procedure of our simple combination of $M_W$ and the transformation between \dalphahad and \amuhvp.

\section*{Supplementary Note 1 - Simple combination of $W$ mass measurements}
\label{app:combination}

We compute a weighted average of $N$ measurements following a standard procedure (see e.g., Ref.~\cite{Zyla:2020zbs}),
\begin{equation}
    \bar x \pm \Delta x = \frac{\sum_{i=1}^N w_i x_i}{\sum_{i=1}^N w_i} \pm \left(\sum_{i=1}^N w_i\right)^{-1/2}
\end{equation}
where $w_i = 1 / \sigma^2_i$. We include the seven measurements ---
LEP~\cite{ALEPH:2013dgf},
LHCb~\cite{LHCb:2021bjt},
ATLAS~\cite{ATLAS:2017rzl},
D0~\cite{D0:2012kms} 92-95 (95/pb) and 02-09 (4.3/fb), CDF~\cite{CDF:2012gpf,CDF:2013dpa,CDF:2022hxs} 88-95 (107/pb) and 02-11 (9.1/fb)
--- avoiding double-counting the CDF data. This resulted in our simple combination
\begin{equation}
    M_W = 80.411 \pm 0.007 \gev.
\end{equation}
This combination remains about $6\sigma$ away from the SM. The chi-squared associated with this estimate,
\begin{equation}
    \chi^2 = \sum_{i=1}^N w_i (x_i - \bar x)^2,
\end{equation}
was $17.7$ with $N - 1 = 6$ degrees of freedom. The associated significance was $2.5\sigma$, found from
\begin{equation}
p = 1 -F_{\chi^2_6}(\chi^2) \quad\text{and}\quad Z = \Phi^{-1}(1 - p)  \end{equation}
where $F_{\chi^2_n}$ is the chi-squared cumulative density function with $n$ degrees of freedom and $\Phi$ is the standard normal cumulative density function.
This result depends on the number and choices of measurements combined. The code to reproduce these calculations is available at \href{https://github.com/andrewfowlie/w\_mass\_combination}{\faGithub}.

We note, however, that a true combination should include careful consideration of correlated systematic errors and expert judgment about unstated or underestimated errors and systematics. For example, if the reduced chi-squared indicates discrepant measurements, the PDG~\cite{Zyla:2020zbs} may inflate the estimated errors, though this does not impact the central value, or decline to combine them (for further discussion see e.g., Ref.~\cite{doi:10.1080/00401706.1972.10488878,Taylor:1982ts,Barlow:2002yb,Jeng:2007is}). In our case, the PDG prescription would inflate the error by about two, if the measurements were combined. This would reduce the discrepancy with the SM to about~$3\sigma$.

\section*{Supplementary Note 2 - The relationship between \dalphahad and \amuhvp}\label{si:hadronic_changes}

Since both \dalphahad and \amuhvp can be extracted from \sigmahad measurements, changes in \sigmahad affect the transformation between \dalphahad and \amuhvp.
On the one hand, one can directly use the experimental data from \sigmahad measurements (the $e^{+}e^{-}$ data) to derive \dalphahad and \amuhvp with the data-driven method as shown in~\autoref{eq:alphahad_sigma_had}
and~\autoref{eq:hvp_sigma_had}
in the main text.
On the other hand, we can extract \dalphahad from EW fits or from estimates of the HVP contributions from the BMW lattice calculation and use that to indirectly indicate possible changes in \sigmahad compared with the experimental measurements. For the latter, we must make assumptions about the energy dependence of \sigmahad and the energy range in which it changes.

As pointed out in Ref.~\cite{Crivellin:2020zul}, we could modify \sigmahad only in the energy ranges:
\begin{align}\label{eq:c1}
&m_{\pi_0} \leq \sqrt{s}\leq 1.937\gev,\\
&m_{\pi_0} \leq \sqrt{s}\leq 11.199\gev\text{ or }\label{eq:c2}\\
&m_{\pi_0} \leq \sqrt{s}\leq \infty,\label{eq:c3}
\end{align}
that is, at low energies, at any moderate energies, or across the entire energy range. The hadronic cross section is unchanged above these thresholds.  We previously only considered the latter possibility~\autoref{eq:c3}; we now consider \autoref{eq:c1}, \autoref{eq:c2} and reconsider the relationship between  \dalphahad and \amuhvp, i.e.\ we follow the procedure introduced in Ref.\ \cite{Crivellin:2020zul}.
We find that using~\autoref{eq:c3} and transforming \dalphahad to \amuhvp (\amuhvp to \dalphahad) results in the most conservative (aggressive) deviation from the $e^{+}e^{-}$ data. However, the low energy range projection~\autoref{eq:c1}  shows the opposite behaviour.
We assume that \sigmahad changes by an overall factor over the range $m_{\pi_0}$ to infinity, \autoref{eq:c3}. This is scenario~(3) in Ref.~\cite{Crivellin:2020zul}. Based on this assumption, we derive \amuhvp from \dalphahad (and vice-versa) with a naive and uniform scaling of the cross-section from the $e^{+}e^{-}$ data~\cite{Aoyama:2020ynm,Keshavarzi:2018mgv}.
In this way, we obtain alternative predictions for \amuhvp that correspond to the $M_W$ measurement under the assumption that no new physics affects the EW fits.

We use the above method to alter the SM prediction for \amu by taking all contributions except for \amuhvp to be those used in Ref.~\cite{Aoyama:2020ynm}. We then determine the deviation between the combined 2021 world average and our predictions ($\delta \amu$), after combining both theoretical and experimental uncertainties. The experimental uncertainty is fixed to $41\times 10^{-11}$~\cite{Muong-2:2021ojo}, but the theoretical uncertainty depends on the way in which \amuhvp was chosen. Each $\delta\amu$ is indicated on the right-hand side of~\autoref{fig:mw_mt_dalphahad}
in the main text
and listed in~\autoref{tab:fits}
in the main text
where we also show how many standard deviations this represents. Finally, we visualize $\delta \amu$ and its tension from~\autoref{tab:fits}
in the main text in~\autoref{fig:muon_g2_new}.

Lastly, to complete our study of the transformation between \dalphahad and \amuhvp, we consider case~\autoref{eq:c1}. Since the BMW collaboration only released the first two bins data ($0 \gev < \sqrt{s}\leq 1\gev$ and $1 \gev < \sqrt{s}\leq\sqrt{10}\gev$) of \dalphahad~\cite{Borsanyi:2020mff}, case~\autoref{eq:c1} may be suggested by BMWc results.
Therefore, we include this case in~\autoref{tab:g2_deviation_fit_low} for readers as a reference. Note we only use the integral breakdown from Ref.~\cite{Keshavarzi:2018mgv,Keshavarzi:2019abf} which provided smaller uncertainties for \amuhvp from the $e^{+}e^{-}$ data in their calculations. Hence, the $\delta \amu$ in~\autoref{tab:g2_deviation_fit_low} would be reduced if the integral breakdown from other references~\cite{Davier:2017zfy,Davier:2019can} was used. A significant caveat to keep in mind is that we don't definitively know whether case~\autoref{eq:c1}, \autoref{eq:c2} or \autoref{eq:c3} should be preferred.

\section*{Supplementary Note 3 - The correlation between $\mathrm{BR}(h\rightarrow\mu\mu)$ and muon $(g-2)$}

As discussed in~\hyperlink{sec:further_constraints}{Further constraints}
in the main text, measurements of the branching ratio $\text{BR}(h\rightarrow \mu\mu)$ \cite{ATLAS:2020fzp,CMS:2020xwi} severely constrain our simplified leptoquark model. Relaxing our simplifications by reintroducing left-handed couplings of the $S_1$ state, $\lambda_L$, substantially reduces the size of the Yukawa couplings needed to explain muon $g-2$. In~\autoref{fig:muon_higgs} we show that this alleviates tension between $g-2$ and $\text{BR}(h\rightarrow \mu\mu)$. The leptoquark model explains muon $g-2$ while satisfying experimental upper limits on $\text{BR}(h\rightarrow \mu\mu)$ for $\lambda_L \gtrsim 10^{-2}$. The branching ratio predictions may be SM-like for $\lambda_L \simeq 5 \times 10^{-2}$. Nevertheless, the model typically predicts deviations from the SM-predictions for $\text{BR}(h\rightarrow \mu\mu)$ that may be observable in the future.

\renewcommand\tablename{Supplementary Table}
\clearpage
\section*{Supplementary Tables}
\begin{table}[!ht]
    \centering
    {\renewcommand{\arraystretch}{1.2}
    \setlength{\tabcolsep}{9pt}
    \centering
    \scalebox{1}{
    \begin{tabular}{lrr}
    \hline
    Parameter & Measured value & Ref.\\
    \hline\hline
    PDG 2021 $M_W$ [GeV] & $80.379(12)$ & \cite{ParticleDataGroup:2020ssz}\\
    CDF 2022 $M_W$ [GeV] & $80.4335(94)$ & \cite{CDF:2022hxs}\\
    \hline\hline\rule{0pt}{3ex}%
    \detaildalphahad & See text\\
    $m_h$ [GeV] & $125.25(17)$ & \cite{ParticleDataGroup:2020ssz}\\
    $m_t$ [GeV]\footnote{$0.5\gev$ theoretical uncertainty is included.} & $172.76(58)$ & \cite{ParticleDataGroup:2020ssz}\\
    $\alpha_s(M_Z)$                        & $0.1179(9)$ & \cite{ParticleDataGroup:2020ssz}\\
    $\Gamma_W$ [GeV]                       & $2.085(42)$ & \cite{ParticleDataGroup:2020ssz}\\
    $\Gamma_Z$ [GeV]                       & $2.4952(23)$ & \cite{ALEPH:2005ab}\\
    $M_Z$ [GeV]                            & $91.1875(21)$ & \cite{ALEPH:2005ab}\\
    $A_{\rm FB}^{0,b}$                     & $0.0992(16)$ & \cite{ALEPH:2005ab}\\
    $A_{\rm FB}^{0,c}$                     & $0.0707(35)$ & \cite{ALEPH:2005ab}\\
    $A_{\rm FB}^{0,\ell}$                  & $0.0171$ & \cite{ALEPH:2005ab}\\
    $A_b$                                  & $0.923(20)$ & \cite{ALEPH:2005ab}\\
    $A_c$                                  & $0.670(27)$ & \cite{ALEPH:2005ab}\\
    $A_\ell(\rm SLD)$                      & $0.1513(21)$ & \cite{ALEPH:2005ab}\\
    $A_\ell(\rm LEP)$                      & $0.1465(33)$ & \cite{ALEPH:2005ab}\\
    $R_b^0$                                & $0.21629(66)$ & \cite{ALEPH:2005ab}\\
    $R_c^0$                                & $0.1721(30)$ & \cite{ALEPH:2005ab}\\
    $R_\ell^0$                             & $20.767(25)$ & \cite{ALEPH:2005ab}\\
    $\sigma_h^0$ [nb]                      & $41.540(37)$ & \cite{ALEPH:2005ab}\\
    $\sin^2\theta_{\rm eff}^{\ell}(Q_{\rm FB})$ & $0.2324(12)$ & \cite{ALEPH:2005ab}\\
    $\sin^2\theta_{\rm eff}^{\ell}({\rm Teva})$ & $0.23148(33)$ & \cite{CDF:2018cnj}\\
    $\overline{m}_c$ [GeV]                 & $1.27(2)$ & \cite{ParticleDataGroup:2020ssz}\\
    $\overline{m}_b$ [GeV]                 & $4.18^{(3)}_{(2)}$ & \cite{ParticleDataGroup:2020ssz}\\
    \hline\hline
    \end{tabular}}
    }
\caption{The measurements included in the global EW fit. Correlations among $(M_Z,\Gamma_Z,\sigma_h^0,R_\ell^0,A_{\rm FB}^{0,\ell})$ and among $(A_{\rm FB}^{0,c},A_{\rm FB}^{0,b},A_c,A_b,R_c^0,R_b^0)$ are also taken into account~\cite{ALEPH:2005ab}.}
\label{tab:fit_input}
\end{table}
\clearpage

\begin{sidewaystable}
    \centering
    \resizebox*{\textwidth}{!}{
    \renewcommand{\arraystretch}{1.25}
    \begin{tabular}{c|c|ccc|ccc|ccc|ccc}
    \hline\hline
    \multicolumn{2}{c|}{$M_W$}& \multicolumn{3}{c|}{Indirect} & \multicolumn{3}{c|}{PDG 2021} & \multicolumn{3}{c|}{CDF 2022} & \multicolumn{3}{c}{Simple Combination}\\
    \multicolumn{2}{c|}{\dalphahad}& BMWc & $e^+e^-$ & Indirect & BMWc & $e^+e^-$ & Indirect & BMWc & $e^+e^-$ & Indirect & BMWc & $e^+e^-$ & Indirect \\
    \hline
    \multirow{2}{*}{Input} & $M_W$ [GeV] & - & - & - & 80.379(12) & 80.379(12) & 80.379(12) & 80.4335(94) & 80.4335(94) & 80.4335(94) & 80.411(7) & 80.411(7) & 80.411(7) \\
    & $\Delta\alpha_{\rm had}^{(5)}(M_Z^2)\times10^4$ & 277.4(1.2) & 276.1(1.1) & - & 277.4(1.2) & 276.1(1.1) & - & 277.4(1.2) & 276.1(1.1) & - & 277.4(1.2) & 276.1(1.1) & - \\
    \hline
    \multirow{7}{*}{Fitted}
    & $\chi^2/\rm dof$ & 16.28/15 & 16.01/15 & 15.89/14 & 19.51/16 & 18.74/16 & 17.59/15 & 65.07/16 & 62.58/16 & 47.19/15 & 52.34/16 & 49.79/16 & 35.48/15 \\
    \cline{2-14}
    & $M_W$ [GeV] & 80.355(6) &  80.357(6) &  80.359(9) & 80.360(6) & 80.361(6) & 80.367(7) & 80.379(5) & 80.380(5) & 80.396(7) & 80.380(5) & 80.381(5) & 80.393(6) \\
    & $\dalphahad\times10^4$ & 277.1(1.2)  & 275.9(1.1) & 274.4(4.4) & 276.8(1.1) & 275.6(1.1) & 271.7(3.8) & 275.6(1.1) & 274.7(1.0) & 260.9(3.6) & 275.6(1.1) & 274.6(1.0) & 262.3(3.4) \\
    \cline{2-14}
    & $\delta a_\mu\times10^{11}$ & - & - & 438(396) & 173(54) & 306(54) & 748(339) & 306(54) & 416(54) & 1997(320) & 306(54) & 416(54) & 1776(301) \\
    & Tension & - & - & $1.1\sigma$ & $3.2\sigma$ & $5.7\sigma$ & $2.2\sigma$ & $5.7\sigma$ & $7.7\sigma$ & $6.2\sigma$ & $5.7\sigma$ & $7.7\sigma$ & $5.9\sigma$ \\
    \cline{2-14}
    & $\delta M_W$ [MeV] & 79(11) & 77(11) & 75(13) & 74(11) & 73(11) & 67(12) & 55(11) & 54(11) & 38(12) & 54(11) & 53(11) & 41(11) \\
    & Tension & $7.2\sigma$ & $7.0\sigma$ & $5.8\sigma$ & $6.7\sigma$ & $6.6\sigma$ & $5.6\sigma$ & $5.0\sigma$ & $4.9\sigma$ & $3.2\sigma$ & $4.9\sigma$ & $4.8\sigma$ & $3.7\sigma$ \\
    \hline\hline
    \end{tabular}
    }
    \caption{SM predictions from EW fits for \dalphahad and $M_W$, and the differences with respect to measurements of muon $g-2$ and the $W$ mass, $\delta \amu$ and $\delta M_W \equiv M_W^{\text{CDF}} - M_W$
    using the low energy projection~\autoref{eq:c1} for the transformation between \dalphahad and \amuhvp.}
    \label{tab:g2_deviation_fit_low}
\end{sidewaystable}

\clearpage

\renewcommand\figurename{{\bf Supplementary Figure}}
\renewcommand\thefigure{{\bf \arabic{figure}}}
\clearpage
\section*{Supplementary Figures}
\begin{figure}[ht]
    \centering
    \includegraphics[width=0.48\textwidth]{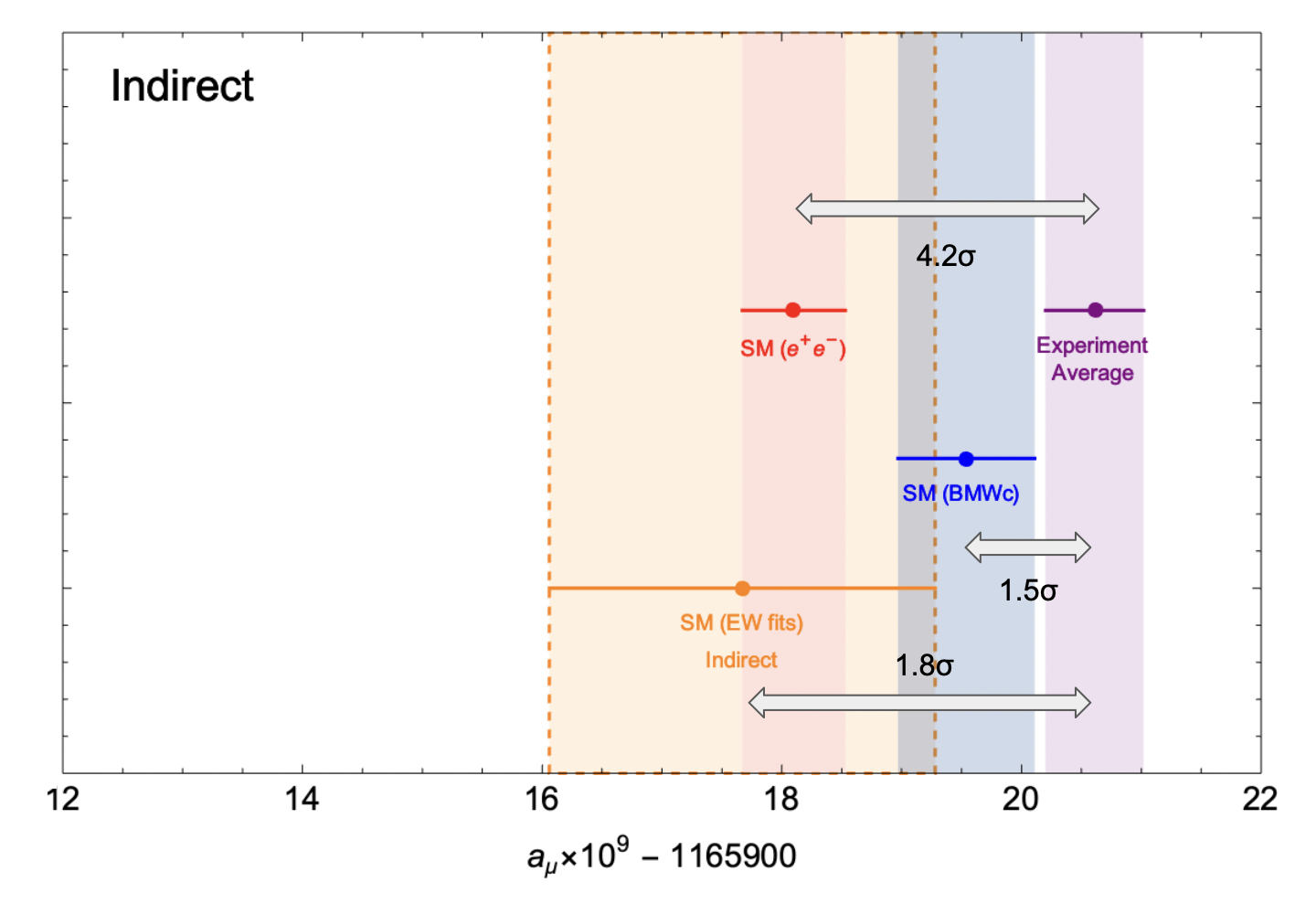}
    \includegraphics[width=0.48\textwidth]{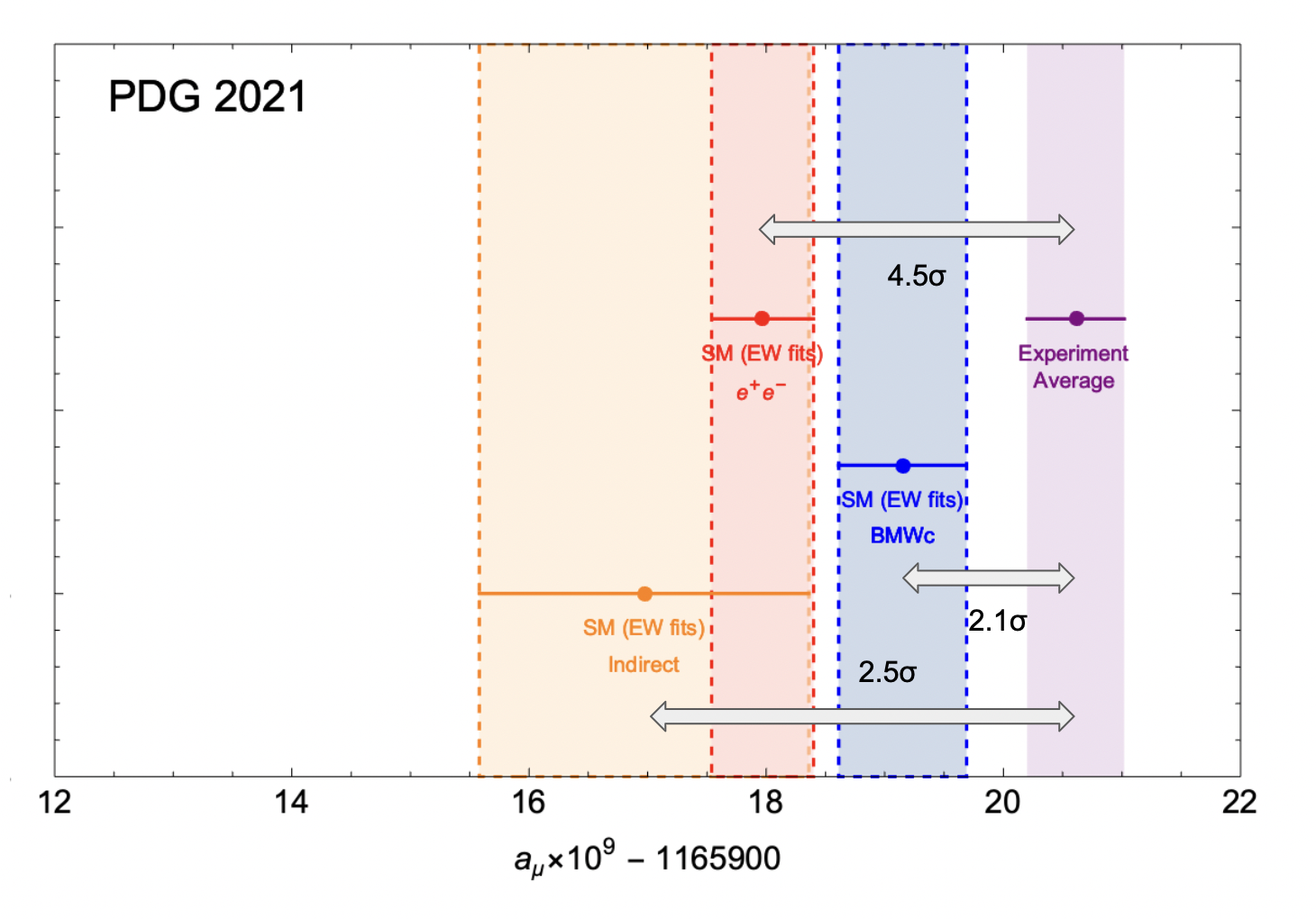}\\
    \includegraphics[width=0.48\textwidth]{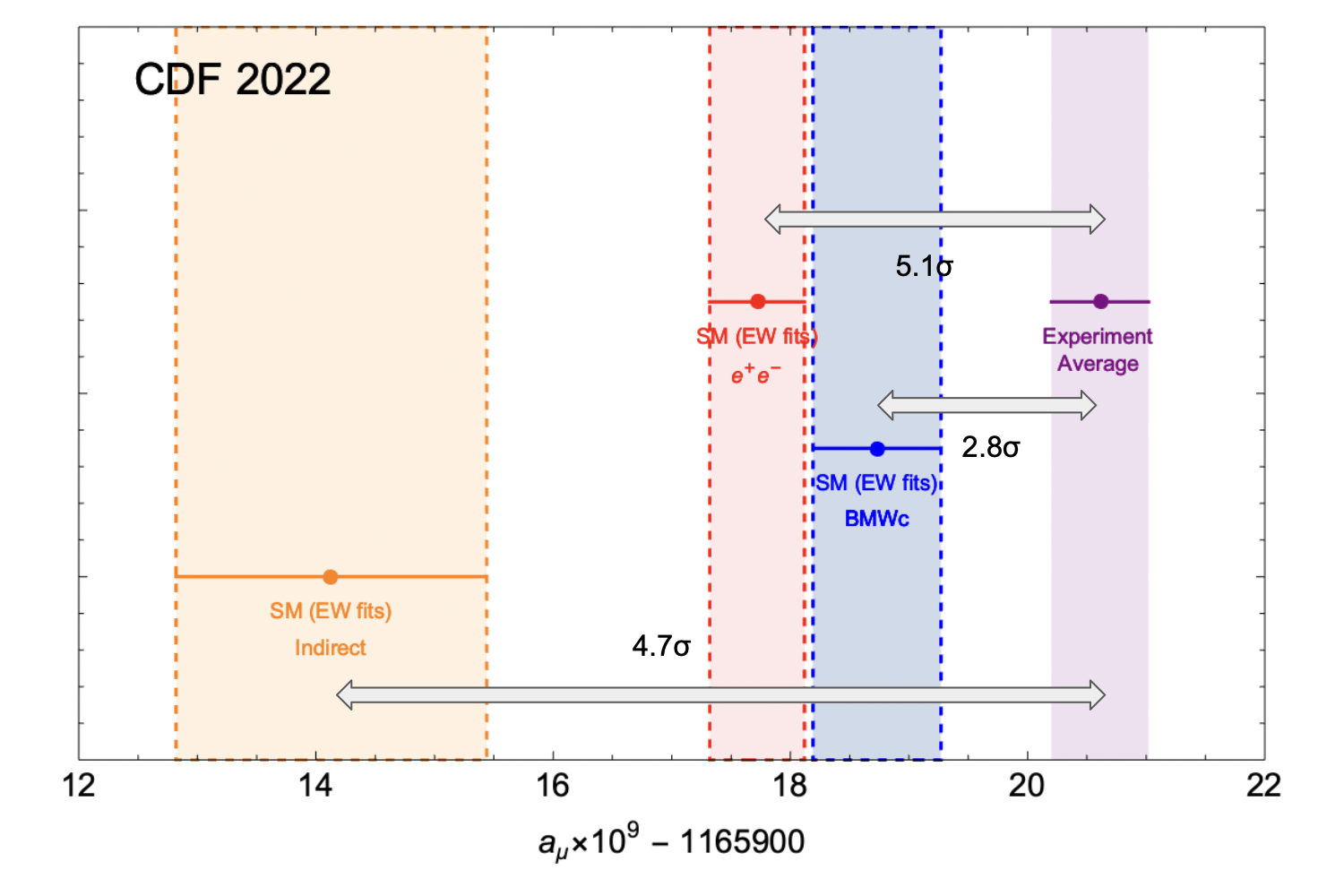}
    \caption{{\bf The $\amu$ in different fitting scenarios and corresponding tensions with experiment average}. The $\amu$ from experiment average of FNAL E989 and BNL E821 (purple), SM predictions from BMWc (blue), $e^{+}e^{-}$ data (red), EW fits w/o \dalphahad (orange-dashed), EW fits with \dalphahad from $e^+e^-$ (red-dashed), EW fits with \dalphahad from BMWc (blue-dashed). The tensions of $\delta \amu$ are also shown for the comparison. The upper-left panel is w/o $M_W$ input, the upper-right and bottom panels are with PDG 2021 and CDF 2022 $M_W$ inputs, respectively.
    }
    \label{fig:muon_g2_new}
\end{figure}

\begin{figure}[ht]
\centering
\includegraphics[width=0.6\textwidth]{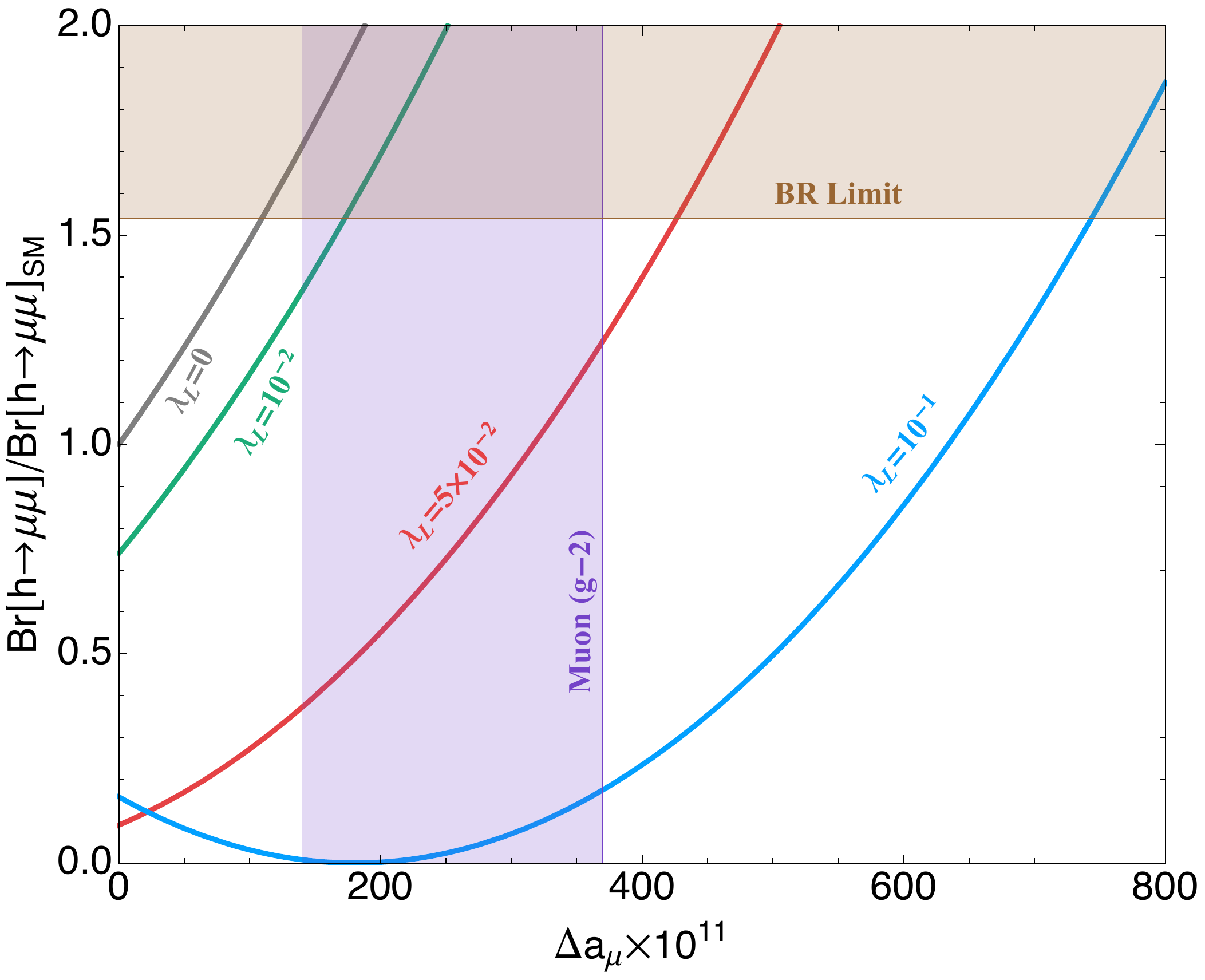}
\caption{{\bf Correlation between the muon $g-2$ and ${\rm BR}(h\to\mu\mu)$.} Correlation between the muon $g-2$ and the branching ratio $\text{BR}(h\rightarrow \mu\mu)$ normalized to its SM value in our leptoquark model. The gray line shows the relationship without the contribution from left-handed couplings ($\lambda_L = 0$), while the green, red, and blue lines correspond to $\lambda_L=10^{-2}$, $5\times 10^{-2}$ and $10^{-1}$, respectively. The mass splitting is fixed at $30\,\mathrm{GeV}$ to be compatible with the $W$ mass measurement. The upper limit on the branching ratio excludes the brown region, whereas the required muon $g-2$ is shown by the purple region.}
\label{fig:muon_higgs}
\end{figure}
\clearpage

\end{document}